\documentclass[11pt]{article}
\usepackage{amsmath,amsfonts,latexsym,graphicx,amssymb}
\usepackage{amsfonts}
\usepackage{bm}
\newcommand{\IS}{\mathbb{I}_{\rm S}}
\newcommand{\IR}{\mathbb{I}_{\rm R}}
\newcommand{\omR}{\omega_{\rm R}}

\newcommand{\ot}{{\,\otimes\,}}

\newcommand{\HR}{{\mathcal{H}_{\rm R}}}
\newcommand{\BH}{{\mathcal{B}(\mathcal{H})}}
\newcommand{\TTH}{{\mathcal{T}(\mathcal{H})}}

\newcommand{\ra}{{\, \rightarrow\, }}
\newcommand{{\Cd}}{{\mathbb{C}^3}}

\def\oper{{\mathchoice{\rm 1\mskip-4mu l}{\rm 1\mskip-4mu l}{\rm 1\mskip-4.5mu l}{\rm 1\mskip-5mu l}}}
\def\<{\langle}
\def\>{\rangle}

\newtheorem{thm}{Theorem}[section]
\newtheorem{proposition}{Proposition}[section]
\newtheorem{definition}{Definition}[section]
\newtheorem{Example}{Example}[section]

\newtheorem{cor}{Corollary}[section]

\begin{document}

\date{}
\title{\textbf{Markovianity criteria for quantum evolution}}

\author{Dariusz  Chru\'sci\'nski and
Andrzej Kossakowski \\ Institute of Physics, Nicolaus Copernicus University,\\
Grudzi\c{a}dzka 5/7, 87--100 Toru\'n, Poland}

\maketitle

\begin{abstract}
 We characterize a class of Markovian dynamics using the concept of divisible dynamical map.  Moreover we provide a family of criteria which can distinguish Markovian and non-Markovian dynamics. These {\em Markovianity criteria} are based on a simple observation that Markovian dynamics implies monotonic behavior of several well known quantities like distinguishability of states, fidelity, relative entropy and genuine entanglement measures.
\end{abstract}

\section{Introduction}

The dynamics of open quantum systems attracts nowadays increasing
attention \cite{Breuer,Weiss,Alicki}.  It is  relevant not only for
the better understanding of quantum theory but it is fundamental in
various modern applications of quantum mechanics. Since the
system-environment interaction causes dissipation, decay and
decoherence it is clear that dynamic of open systems is fundamental
in modern quantum technologies, such as quantum communication,
cryptography and computation \cite{QIT}.

The usual approach to the dynamics of an open quantum system
consists in applying an appropriate Markovian approximation, that leads to the
following local master equation
\begin{equation}
\label{M-0}
    \dot{\rho}_t = L\, \rho_t\ ,
\end{equation}
where $\rho_t$ is the density matrix of the system investigated and
$L$ the time-independent generator of the dynamical semigroup. It is well known that under certain conditions on $L$ \cite{GKS,Lindblad}
the corresponding dynamics $\rho \ra \rho_t := \Lambda_t\rho $ gives rise to completely positive and trace preserving map $\Lambda_t$
\cite{Alicki,Breuer} (one usually calls $\Lambda_t$ a {\em dynamical map}).
The characteristic feature of Markovian approximation leading to dynamical semigroup $\Lambda_t = e^{tL}$  is that it neglects all
memory effects caused by the nontrivial interaction of the system with an external world.
However, recent theoretical studies and technological progress call for more refine approach based on
non-Markovian evolution.

Non-Markovian systems appear in many branches of physics, such as
quantum optics \cite{Breuer,Gardiner}, solid state physics
\cite{solid}, quantum chemistry \cite{Plenio-K}, and quantum
information processing \cite{Aharonov}. Since non-Markovian dynamics
modifies monotonic decay of quantum coherence it turns out that
when applied to composite systems it may protect quantum
entanglement for longer time than standard Markovian evolution
\cite{Saverio}. In particular it may protect the system against the
sudden death of entanglement \cite{DEATH}. It is therefore not
surprising that non-Markovian dynamics was intensively studied
during last years \cite{NM}--\cite{PRL}.

The standard approach to the dynamics of open system uses the
Nakajima-Zwanzig projection operator technique \cite{NZ} which shows
that under fairly general conditions, the master equation for the
reduced system density matrix $\rho_t$ takes the form of the following
non-local equation
\begin{equation}\label{NM-0}
    \dot{\rho}_t = \int_0^t \mathcal{K}_{t-\tau}\, \rho_\tau \, d\tau\ ,
\end{equation}
in which quantum memory effects are taken into account through the
introduction of the memory kernel $\mathcal{K}_t$: this simply
means that the rate of change of the state $\rho_t$ at time $t$
depends on its history (starting at $t=0$). It should be stresses that there is an perfectly equivalent approach [usually called time convolutionless  (TCL)] \cite{Breuer,TCL} which describes quantum dynamics by local in time equation
\begin{equation}\label{Local-0}
    \dot{\rho}_t = L_t \rho_t \ ,
\end{equation}
where $L_t$ denotes local generator. It is clear that if $L_t=L$ does not depend on time and $\mathcal{K}_t = 2 \delta(t) L$, then both (\ref{NM-0}) and (\ref{Local-0}) reduce to the standard master equation (\ref{M-0}). These equations  provide, therefore,  natural generalization of the standard Markovian semigroup. One of the open problems in this theory is to characterize properties of $\mathcal{K}_t$ and $L_t$ which guarantee that the corresponding solution $\rho_t = \Lambda_t \rho$ defines legitimate dynamical map $\Lambda_t$.

Interestingly,  the concept of (non)Markovianity is not
uniquely defined. One approach is based on the idea of the
composition law which is essentially equivalent to the idea of
divisibility \cite{Wolf2}. This approach was used recently \cite{RHP} to construct the corresponding
measure of non-Markovianity. A different approach is presented in \cite{BLP} where non-Markovian dynamics corresponds to
a time evolution for the open system characterized by a temporary
flow of information from the environment back into the system and
manifests itself as an increase in the distinguishability of pairs
of evolving quantum states. The aim of this paper is to characterize a class of Markovian dynamics using the concept of divisible dynamical map and to provide a family of criteria which can distinguish Markovian and non-Markovian dynamics, i.e.  these criteria are satisfied if the dynamics is Markovian and hence their violation is a clear sign of non-Markovianity.

The paper is organized as follows: in the next Section we recall  the standard reduced dynamics of an open system. We stress that the problem of characterizing the properties of $\mathcal{K}_t$ and/or $L_t$ is in general untractable. It considerably simplifies in the case of commutative dynamics, i.e if the dynamical map $\Lambda_t$ commutes in different times $[\Lambda_t,\Lambda_u]=0$ for arbitrary $t,u \geq 0$ (see Section \ref{COMM}). In Section \ref{MAR} we characterize Markovian dynamics using the concept of divisible dynamical map. Interestingly, we provide a simple method which enables one to go beyond Markovian dynamics fully controlling the corresponding local generator. This method is illustrated by pure decoherence.   Necessary criteria for Markovianity are discussed in Section \ref{CRI}. Final conclusions are collected in Section \ref{CON}.

\section{Reduced dynamics of an open system}

Consider an $N$-level quantum system living in $\mathcal{H}$ coupled to a reservoir with the corresponding (usually infinite dimensional) Hilbert space $\HR$.  Throughout the paper we use standard notation: $\BH$ and $\TTH$ denote the class of bounded and trace class operataors in $\mathcal{H}$, respectively.
Let $H$ denotes the Hamiltonian of the total composed system living in $\mathcal{H}\ot \HR$ and $\omega$ be a fixed state of the reservoir. One defines the reduced dynamics $\Lambda_t : \TTH \ra \TTH$ by the following formula
\begin{equation}\label{RED}
    \Lambda_t \rho := {\rm Tr}_{\rm R} \, \Big[ e^{-i\mathbf{H}t}\,(\rho \ot \omR  )\, e^{i\mathbf{H}t} \Big] \ ,
\end{equation}
where ${\rm Tr}_{\rm R}$ denotes the partial trace over the reservoir degrees of freedom.  Note, that $\Lambda_t$ is completely positive and trace preserving for all $t \geq 0$ and it satisfies $\Lambda_0 = \oper$. Therefore, it provides a legitimate quantum evolution of the system living in $\mathcal{H}$. Actually, it is well known that any legitimate $\Lambda_t$ may be defined as a reduced dynamics for appropriate $\HR$ and the total Hamiltonian $\mathbf{H}$. The standard Nakajima-Zwanzig projection operator technique \cite{NZ,Breuer}  shows that reduced dynamics $\Lambda_t$ satisfies the following non-local equation
\begin{equation}\label{NZ}
    \dot{\Lambda}_t = \int_0^t \mathcal{K}_{t-\tau}\, \Lambda_\tau \, d\tau\ ,\ \ \ \ \Lambda_0 = \oper\ .
\end{equation}
The memory kernel $\mathcal{K}_t$ encodes all dynamical properties of the system and depends upon the total Hamiltonian of the ``system + reservoir'' and the reservoir reference state $\omR$. This equation is exact but in general very difficult to analyze. This is due to the fact that the memory kernel $\mathcal{K}_t$ depends upon all reservoir correlation functions. To simplify analysis one usually tries to perform suitable Markovian approximation to neglect all unwanted memory effects. The validity of such approximation is based on the existence of two characteristic time scales: the characteristic time $\tau_S$ of variation of $\rho_t$ and the decay time $\tau_R$ of the reservoir correlation functions. Markovian approximation assumes that $\tau_S \gg \tau_R$. Basically there are two ways of rigorous treatment of the limit $\tau_S/\tau_R \rightarrow \infty$. One assumes that $\omR$ is invariant under the free evolution of the reservoir. Representing the Hamiltonian $\mathbf{H}$ as
\begin{equation}\label{}
    \mathbf{H} = H_{\rm S} \ot \IR + \IS \ot H_{\rm R} + \lambda H_{\rm int}\ ,
\end{equation}
one performs the weak coupling limit $\lambda \rightarrow 0$ with rescaled time $\tau = \lambda^2 t$. In this scheme $\tau_R$ remains constant, while $\tau_S \rightarrow \infty$. This approach was analyzed in great details by Davies \cite{Davies,Davies1} (see also \cite{Dumcke}).

On the other hand in the singular coupling limit one has $\tau_R \rightarrow 0$. It is achieved by considering the following Hamiltonian
\begin{equation}\label{}
    \mathbf{H} = H_{\rm S} \ot \IR + \epsilon^{-2} \IS \ot H_{\rm R} + \epsilon^{-1} H_{\rm int}\ ,
\end{equation}
and performing the limit $\epsilon \rightarrow 0$ \cite{Palmer}. As a result reservoir correlation functions become $\delta$ function.
In both scenarios the limiting dynamics is governed by the well known master equation
\begin{equation}\label{M1}
  \dot{\Lambda}_t = L \Lambda_t\ ,     \ \ \ \Lambda_0=\oper\ ,
\end{equation}
where $\oper$ denotes the identity map, and the Markovian generator  $L$ is given by
\begin{equation}\label{GKSL}
    L \, \rho = -i[H,\rho] + \frac 12 \sum_\alpha \left( [V_\alpha,\rho
    V_\alpha^\dagger] + [V_\alpha\rho,V_\alpha^\dagger] \right) \ .
\end{equation}
In what follows we call $L$ represented by (\ref{GKSL}) {\em GKSL generator}.
In the above formula $H$ represents the effective system Hamiltonian and
$\{V_\alpha \}$ is the collection of arbitrary operators encoding
the interaction between system and the environment. Equation (\ref{M1}) gives rise to Markovian semigroup $\Lambda_t = e^{L t}$ satisfying the following homogeneous composition law
\begin{equation}\label{}
    \Lambda_t \, \Lambda_u = \Lambda_{t+u}\ ,
\end{equation}
for any $t,u \geq 0$.

It should be stressed that one obtains Markovian master equation (\ref{M1}) from the general Nakajima-Zwanzig equation (\ref{NZ}) only if the total Hamiltonian $\mathbf{H}$ enables one to perform suitable Markovian approximation and hence it covers only limited number of physically interesting systems. In general Markovian approximation is not suitable and one has to deal with much more involved non-local equation (\ref{NZ}). One of the main problems is to characterize the properties of the corresponding memory kernel $\mathcal{K}_t$ which guaranties that the corresponding solution $\Lambda_t$ represents legitimate dynamical map.

Note, that instead of non-local equation (\ref{NZ}) one may equivalently describes dynamics using local equation. This approach (usually called time-convolusionless  \cite{Breuer}) leads to the following local in time equation
\begin{equation}\label{Local}
    \dot{\Lambda}_t = L_t \Lambda_t\ ,     \ \ \ \Lambda_0=\oper\ ,
\end{equation}
with time-dependent local generator $L_t$. We stress that these two approaches are equivalent.  Assuming that $\Lambda_t$ is differentiable it always satisfies local in time master equation.  Indeed, formally one has $\dot{\Lambda}_t = \dot{\Lambda}_t\Lambda_t^{-1}\Lambda_t = L_t \Lambda_t$, where we assumed the existence of the inverse map $\Lambda_t^{-1}$. Note, that the inverse, even if it exists, needs not be completely positive. Again, one would like to perform  characterization of time-dependent generators $L_t$ giving rise to legitimate dynamical maps $\Lambda_t$.  This problem seems to be untractable in full generality. Note, that formal solution to (\ref{Local}) has the following form
\begin{equation}\label{T-exp}
    \Lambda_t = {\rm T} \, \exp\left( \int_0^t L_u du \right) \ ,
\end{equation}
where `T' stands for chronological product. It is clear that the above expression has rather a formal meaning.

\section{Commutative class of dynamical maps} \label{COMM}

As we already stressed the general solution to the local in time master equation (\ref{Local}) has only a formal meaning and in general we do not control the properties of $L_t$ which guarantee that T-product exponential formula ${\rm T} \exp( \int_0^t L_u du)$ defines dynamical map. Note, however, that if $L_t$ defines a commutative family of generators, that is
\begin{equation}\label{}
    [L_t,L_u]=0\ ,
\end{equation}
for any $t,u \geq 0$, the formula (\ref{T-exp}) considerably simplifies: the `T' product drops out and the solution is fully controlled by the integral $\int_0^t L_u du$.

\begin{thm}
If $L_t$ defines a commutative family, then $L_t$ is a legitimate generator of a quantum dynamical map if and only if $\int_0^t L_u du $ defines a legitimate GKSL generator for all $t\geq 0$.
\end{thm}

\begin{Example} {\em As an example of commutative dynamics consider the following evolution of the qubit
\begin{equation}\label{L-2}
    L_t \rho = - \frac{i \omega_t}{2} [\sigma_z,\rho] + \frac{\gamma_t}{2} (\sigma_z \rho \sigma_z - \rho)\ ,
\end{equation}
where $\omega_t,\, \gamma_t : \mathbb{R}^+ \ra \mathbb{R}$. The corresponding density matrix evolves as follows
\begin{equation}\label{}
    \rho_t =
    \left( \begin{array}{cc} \rho_{00} & \rho_{01} e^{(-i \Omega_t - \Gamma_t)} \\ \rho_{10}e^{(i \Omega_t - \Gamma_t)} & \rho_{11} \end{array} \right)\ ,
\end{equation}
where
$$   \Omega_t = \int_0^t \omega_u du\ , \ \ \  \Gamma_t = \int_0^t \gamma_u du\ . $$
$L_t$ gives rise to legitimate quantum evolution if and only if $\Gamma_t \geq 0$ for all $t\geq 0$ and hence the evolution corresponds to simple decoherence. The corresponding dynamics is Markovian iff $\gamma_t \geq 0$.

This example may be easily generalized for $d$-level system. Consider the following class of generators: let $\lambda= e^{2\pi i/d}$ and define
\begin{equation}\label{}
    V_\alpha = \sum_{\beta=0}^{d-1} \lambda^{\alpha\beta} P_\beta\ ,
\end{equation}
where $P_\beta = |e_\beta\>\<e_\beta|$ and $\{e_0,\ldots,e_{d-1}\}$ stands for an arbitrary orthonormal basis in $\mathbb{C}^d$. Now, let $c_{\alpha\beta}(t)$ be a time-dependent hermitian matrix and define
\begin{equation}\label{}
    L_t \rho = -i[H_t,\rho] + \sum_{k,l=1}^{d-1} c_{kl}(t) \left( [V_k,\rho
    V_l^\dagger] + [V_k\rho,V_l^\dagger] \right) \ ,
\end{equation}
where the time-dependent Hamiltonian $H_t$ reads as follows
$$ H_t = \sum_{k=1}^{d-1} (h_k(t) V_k + \overline{h}_k(t) V^\dagger_k)\ , $$
where $h_k: \mathbb{R}^+ \ra \mathbb{C}$.  $L_t$ generates legitimate dynamics if and only if
the matrix
$$ C_{kl}(t) = \int_0^t c_{kl}(u) du \, $$
is positive definite. Dynamics is Markovian if and only if $c_{kl}(t)$ is itself positive definite.

}
\end{Example}

\section{Markovian dynamics}  \label{MAR}

In this section we characterize important class of dynamical maps representing Markovian evolution. It is important to clarify this issues since there are few definitions  used in the literature recently. We call a dynamical map $\Lambda_t$ {\em divisible} if for any $t \geq s \geq 0$ one has the following decomposition
\begin{equation}\label{}
    \Lambda_t = V_{t,s}\, \Lambda_s \ ,
\end{equation}
with completely positive propagator $V_{t,s}$. Note, that $V_{t,s}$ satisfies inhomogeneous composition law
\begin{equation}\label{}
    V_{t,s} V_{s,u} = V_{t,u}\ ,
\end{equation}
for any $t\geq s \geq u$. In this paper following \cite{RHP} we accept the following
\begin{definition}
Dynamical map $\Lambda_t$ corresponds to Markovian evolution if and only if it is divisible.
\end{definition}
Interestingly, the property of being Markovian (or divisible) is fully characterized in terms of the local generator $L_t$. Note, that if $\Lambda_t$ satisfies (\ref{Local}) then $V_{t,s}$ satisfies
\begin{equation}\label{Local-V}
    \partial_t\, V_{t,s} = L_t V_{t,s}\ ,     \ \ \ V_{s,s}=\oper\ ,
\end{equation}
and the corresponding solution reads
\begin{equation}\label{}
    V_{t,s} = {\rm T} \, \exp\left( \int_s^t L_u du \right) \ .
\end{equation}
The central result consists in the following

\begin{thm}
The map $\Lambda_t$ is divisible if and only if $L_t$ has the GKSL form for all $t$.
\end{thm}

\noindent Proof: (the proof goes similarly as for time-independent case (cf. \cite{Breuer,Alicki}) Assume that $\Lambda_t$ is divisible, that is, $V_{t,s}$ satisfies (\ref{Local-V}). One has
\begin{equation}\label{}
    L_t = \lim_{\epsilon\rightarrow 0} \frac{V_{t+\epsilon,t} - \oper}{\epsilon}\ ,
\end{equation}
for any $t$. Now, to compute $L_t$ let $F_\alpha$ ($\alpha = 0,1,\ldots,d^2-1$) denotes an orthonormal basis in $M_d(\mathbb{C})$ such that $F_0 = \mathbb{I}_d/\sqrt{d}$. Now, since $V_{t,s}$ is completely positive one has the corresponding Kraus representation
\begin{equation}\label{}
    V_{t,s}\, \rho = \sum_{\alpha,\beta=0}^{d^2-1} c_{\alpha\beta}(t,s) \, F_\alpha \rho F_\beta^\dagger\ ,
\end{equation}
where the matrix $c_{\alpha\beta}(t,s)$ is positive definite for all $t \geq s\,$, and $V_{t,t}=\oper$ implies
$\, c_{\alpha\beta}(t,t)=d\, \delta_{\alpha 0} \delta_{\beta 0}  \,$
for all $\alpha, \beta = 0,1,\ldots,d^2-1\,$. One finds
\begin{eqnarray}
   L_t\, \rho &=& \lim_{\epsilon\rightarrow 0} \left\{ \frac{c_{00}(t+\epsilon,t) - d}{\epsilon} \, F_0\rho F_0 + \sum_{k=1}^{d^2-1}  \frac{c_{0k}(t+\epsilon,t)}{\epsilon} \, F_0 \rho F_k^\dagger \right. \nonumber \\
   &+& \left.  \sum_{k=1}^{d^2-1}  \frac{\overline{c}_{0k}(t+\epsilon,t)}{\epsilon} \, F_k \rho F_0 + \sum_{k,l=1}^{d^2-1}  \frac{c_{kl}(t+\epsilon,t)}{\epsilon} \, F_k \rho F_l^\dagger \right\} \ .
\end{eqnarray}
Let us introduce
\begin{equation*}\label{}
    a_{00}(t) = \lim_{\epsilon\rightarrow 0} \frac{c_{00}(t+\epsilon,t) - d}{\epsilon}\ , \ \ \
    a_{0k}(t) = \lim_{\epsilon\rightarrow 0} \frac{c_{0k}(t+\epsilon,t)}{\epsilon}\ , \ \ \
    a_{kl}(t) = \lim_{\epsilon\rightarrow 0} \frac{c_{kl}(t+\epsilon,t)}{\epsilon}\ . \ \ \
\end{equation*}
Note, that the Hermitian time-dependent matrix $a_{kl}(t)$ is positive definite. Moreover, let
\begin{equation*}
    G_t = \frac{1}{2d}\, a_{00}(t) \mathbb{I}_d + \frac 12 (A_t + A_t^\dagger) \ , \ \ \ \
    H_t = \frac{1}{2i} (A_t - A^\dagger_t)\ ,
\end{equation*}
with $A_t$ defined by
\begin{equation*}
    A_t = \frac{1}{\sqrt{d}}\, \sum_{k=1}^{d^2-1} a_{0k}(t) F_k \ .
\end{equation*}
Finally, one obtains the following formula for the local generator
\begin{equation}\label{}
    L_t\rho =  -i[H_t,\rho] + \{G_t,\rho\} +  \sum_{k,l=1}^{d^2-1} a_{kl}(t) \, F_k \rho F_l^\dagger\ .
\end{equation}
Taking into account that $\Lambda_t$ is trace preserving one has ${\rm Tr}(L_t \rho)=0$ for all $\rho$ which implies
\begin{equation}\label{Gt}
    G_t = - \frac 12 \sum_{k,l=1}^{d^2-1} a_{kl}(t)\, F^\dagger_l F_k\  \ ,
\end{equation}
and hence
\begin{equation}\label{ST}
    L_t\rho = -i[H_t,\rho] + \sum_{k,l=1}^{d^2-1} a_{kl}(t) \left( F_k \rho F_l^\dagger - \frac 12 \{ F_l^\dagger F_k,\rho\} \right) \ ,
\end{equation}
reproduces the standard GKSL form of $L_t$ (recall that $a_{kl}(t)$ is positive definite).

Assume now, that $L_t$ is defined by (\ref{ST}). It may be rewritten as follows
\begin{equation}\label{Phi-Psi}
    L_t = \Phi_t - \Psi_t\ ,
\end{equation}
where $\Phi_t$ is a family of completely positive maps
\begin{equation}\label{}
    \Phi_t \rho = \sum_{k,l=1}^{d^2-1} a_{kl}(t)  F_k \rho F_l^\dagger\ ,
\end{equation}
and
\begin{equation}\label{}
    \Psi_t \rho = C_t \rho - \rho C^\dagger_t \ ,
\end{equation}
where
\begin{equation}\label{}
    C_t = i H_t +  G_t\ .
\end{equation}
Actually, due to (\ref{Gt}) one has $G_t = - \frac 12 \Phi^\#_t \mathbb{I}$.
Note, that by construction ${\rm Tr}(L_t \rho)=0$, and hence the corresponding solution $V_{t,s}$
is trace preserving. It remains to show that $V_{t,s}$ is completely positive for all $t \geq s$.
Consider the following equation
\begin{equation}\label{}
     \partial_t\, N_{t,s} = -\Psi_t N_{t,s}\ ,     \ \ \ N_{s,s}=\oper\ .
\end{equation}
One easily finds
\begin{equation}\label{N-ts}
    N_{t,s} \, \rho = X_{t,s}\, \rho\, X_{t,s}^\dagger\ ,
\end{equation}
where $X_{t,s}$ itself satisfies $\, \partial_t X_{t,s} = C_t X_{t,s}\,$ and hence
\begin{equation}\label{}
    X_{t,s} = {\rm T} \exp\left( \int_s^t C_u du \right)\ .
\end{equation}
It is clear that $N_{t,s}$ is completely positive. Moreover, it is invertible and the inverse $N_{t,s}^{-1}$ reads
\begin{equation}\label{}
    N_{t,s}^{-1}\rho = Y_{t,s}\, \rho\, Y_{t,s}^\dagger\ ,
\end{equation}
where
\begin{equation}\label{}
    Y_{t,s} = {\rm T}_0 \exp\left( -\int_s^t C_u du \right)\ ,
\end{equation}
with ${\rm T}_0$ denoting  anti-chronological operator. Hence  $ N_{t,s}^{-1}$ is completely positive as well.  To solve original equation (\ref{Local-V}) let us pass to the ``interaction'' picture and define
\begin{equation}\label{VNV}
 V_{t,s} = N_{t,s} V_{t,s}^{({\rm int})} \ .
\end{equation}
One finds
\begin{equation}\label{L12}
     \partial_t\, V_{t,s}^{({\rm int})} = \Phi_{t,s}^{({\rm int})} V_{t,s}^{({\rm int})}\ ,     \ \ \ V_{s,s}^{({\rm int})}=\oper\ ,
\end{equation}
where
\begin{equation}\label{NPN}
    \Phi_{t,s}^{({\rm int})} = N_{t,s}^{-1} \circ \Phi_t \circ N_{t,s}\ .
\end{equation}
The above formula shows that $\Phi_{t,s}^{({\rm int})}$ is completely positive being the composition of three completely positive maps: $N_{t,s}$, $\Phi_t$ and $N_{t,s}^{-1}$.  One easily solves (\ref{L12}) and gets
\begin{equation}\label{}
     V_{t,s}^{({\rm int})} = {\rm T}\, \exp\left( \int_s^t \Phi_{u,s}^{({\rm int})} \, du \right) \ .
\end{equation}
It is therefore clear that $V_{t,s}^{({\rm int})}$ can be represented as the following series
\begin{equation}\label{V-CP}
    V_{t,s}^{({\rm int})} = \oper + \int_s^{t} dt_1\, \Phi_{t_1,s}^{({\rm int})} + \int_s^t dt_1 \int_s^{t_1} dt_2\,   \Phi_{t_1,s}^{({\rm int})} \circ \Phi_{t_2,s}^{({\rm int})} + \ldots\ .
\end{equation}
It shows that $V_{t,s}^{({\rm int})}$ is completely positive being a sum of completely positive maps.  Hence, taking into account formula (\ref{VNV}) it finally shows that $V_{t,s}$ is completely positive. \hfill $\Box$

\vspace{.3cm}

It is clear from (\ref{V-CP}) that complete positivity of $\Phi_{t,s}^{({\rm int})}$ is sufficient for complete positivity of $V_{t,s}^{({\rm int})}$. Note, that formula (\ref{NPN}) implies
\begin{equation}\label{NPN-d}
     \Phi_{t}^\# = N_{t,s}^\# \circ \Phi_{t,s}^{({\rm int})\#} \circ N_{t,s}^{\# -1}\ ,
\end{equation}
where $\Lambda^\# : \BH \ra \BH$ denotes a dual map defined by
\begin{equation}\label{}
    {\rm Tr}(\rho \cdot \Lambda^\# a) =  {\rm Tr}(a \cdot \Lambda\rho  )\ ,
\end{equation}
for all $\rho \in \TTH$ and $a \in \BH$. Now, if $\Phi_t^{({\rm int})}$ is completely positive then formula (\ref{NPN-d}) proves that $\Phi_t$ is completely positive as well. Hence, complete positivity of $\Phi_t$ cannot be relaxed.

Let us observe, that a class of Markovian evolution may be easily generalized as follows: we are looking for the solution of (\ref{Local}) with $L_t$ represented as in (\ref{Phi-Psi}). Our aim is find $\Phi_t$ and $\Psi_t$ such that $L_t$ defines a legitimate generator. Suppose, that we are given a completely positive map $N_t$ satisfying $N_0=\oper$. Suppose that $N_t$ is not trace preserving (if it were  it represents legitimate dynamics and we are done). Let us define $\Psi_t$ as a local in time generator such that
\begin{equation}\label{}
    \dot{N}_t = - \Psi_t N_t\ , \ \ \ N_0=\oper\ .
\end{equation}
One obviously has $\Psi_t = - \dot{N}_t N_t^{-1}$. Now, the question is: does there exit $\Phi_t$ such that $L_t = \Phi_t - \Psi_t$ provides legitimate generator? Note, that the solution of (\ref{Local}) has the following form $\Lambda_t = N_t \Lambda_{t}^{({\rm int})}$, where $\Lambda_{t}^{({\rm int})}$ is defined in terms of the following series
\begin{equation}\label{}
    \Lambda_{t}^{({\rm int})} = \oper + \int_0^{t} dt_1\, \Phi_{t_1}^{({\rm int})} + \int_0^t dt_1 \int_0^{t_1} dt_2\,   \Phi_{t_1}^{({\rm int})} \circ \Phi_{t_2}^{({\rm int})} + \ldots\ ,
\end{equation}
with $ \Phi_{t}^{({\rm int})} = N_{t}^{-1} \circ \Phi_t \circ N_{t}$. It is clear that if $\Phi_{t}^{({\rm int})}$ is completely positive so is $\Lambda_t$. Now,
\begin{equation}\label{Theta}
    \Phi_{t} = N_{t} \circ \Phi_t^{({\rm int})} \circ N_{t}^{-1} = \Theta_t N_{t}^{-1}\ ,
\end{equation}
with $\Theta_t := N_{t}  \Phi_t^{({\rm int})}$ being a  completely positive map. Note, that $\Phi_t$ needs not be completely positive. However, if $N_t^{-1}$ is completely positive then necessarily $\Phi_t$ is completely positive as well. Hence, if $\Phi_t$ is constructed by (\ref{Theta}), then $\Lambda_t$ is completely positive. It remain to check for trace preservation. Note, that $\Lambda_t$ is trace preserving if $L^\#_t \mathbb{I} =0$. One has
\begin{equation}\label{}
    L_t^\# \mathbb{I} = N_t^{-1 \#} ( \Theta^\#_t + \dot{N}^\#_t ) \mathbb{I}\ ,
\end{equation}
and hence if
\begin{equation}\label{Theta-N}
    \Theta^\#_t \mathbb{I} + \dot{N}^\#_t  \mathbb{I} = 0 \ ,
\end{equation}
then $\Lambda_t$ defines legitimate dynamical map. It is therefore clear that if
\begin{equation}\label{dot-N}
    - \dot{N}^\#_t  \mathbb{I} \geq 0 \ ,
\end{equation}
then one can always find completely positive $\Theta_t$ such that normalization condition (\ref{Theta-N}) holds. Clearly, the choice of $\Theta_t$ is highly non unique. If $\Theta_t$ satisfies (\ref{Theta-N}) and $M_t$ is an arbitrary family of quantum channels, then the following `gauge transformation' $\Theta_t \ra \Theta_t^M := M_t \Theta_t$ gives rise to another admissible $\Theta_t^M$ satisfying (\ref{Theta-N}).

\begin{proposition}
If $N_t$ with $N_0=\oper$ is a family of completely positive map satisfying (\ref{dot-N}), then there exists completely positive $\Theta_t$ satisfying (\ref{Theta-N}) such that $L_t = (\Theta_t + \dot{N}_t)N_{t}^{-1}$ gives rise to the legitimate local generator.
\end{proposition}
Actually, condition (\ref{dot-N}) is always satisfied for the Markovian evolution. Indeed, taking into account (\ref{N-ts}) one finds
\begin{equation}\label{}
     \partial_t{N}_{t,s}^\#  \mathbb{I}  =  (\partial_t X^\dagger_{t,s}) X_{t,s} + X^\dagger_{t,s}(\partial_t X_{t,s}) \ ,
\end{equation}
and hence using $\, \partial_t X_{t,s} = C_t X_{t,s}\,$
one finds
\begin{equation}\label{dot-NN}
     \partial_t{N}_{t,s}^\#  \mathbb{I}  =  X^\dagger_{t,s}\left[ C_t + C_t^\dagger \right] X_{t,s}  = 2 X^\dagger_{t,s} G_t X_{t,s} \ ,
\end{equation}
which shows that  $\partial_t{N}_{t,s}^\#  \mathbb{I} \leq 0$ due to $G_t \leq 0$.  Therefore, presented method generalizes Markovian generator keeping $N_t$ completely positive and satisfying (\ref{dot-N}) but admitting $\Phi_t$ to be not completely positive.
It should be stressed that this construction provides a local analog of semi-Markov dynamics constructed recently in \cite{semi}.

\begin{Example} Consider the following family of completely positive maps
\begin{equation}\label{}
    N_t\rho = \sum_{k,l=1}^N n_{kl}(t)\, e_{kk} \, \rho\,  e_{ll}\ ,
\end{equation}
where $e_{ij} = |i\>\<j|$, and the matrix $n_{kl}(t)$ is positive definite with $n_{kl}(0)=1$.  One easily finds for $\Psi_t$
\begin{equation}\label{}
    \Psi_t\rho :=  - \dot{N}_t N_t^{-1}\rho = - \sum_{k,l=1}^N \frac{\dot{n}_{kl}(t)}{n_{kl}(t)} \, e_{kk}\, \rho\, e_{ll}\ .
\end{equation}
Note, that condition (\ref{dot-N}) is equivalent to the following condition for the diagonal elements
\begin{equation}\label{}
    \dot{n}_{kk}(t) \leq 0\ ,
\end{equation}
which implies $n_{kk}(t) \leq n_{kk}(0)=1$.
Now, let $\Theta_t$ be a family of completely positive maps
\begin{equation}\label{}
\Theta_t\rho = \sum_{k,l=1}^N \theta_{kl}(t)\, e_{kk} \, \rho\,  e_{ll}\ ,
\end{equation}
where the matrix $\theta_{kl}(t)$ is positive definite. Normalization condition (\ref{Theta-N}) shows that the diagonal elements of the matrix $\theta_{kl}(t)$ are uniquely determined by
 $   \theta_{kk}(t) = - \dot{n}_{kk}(t)\,$.
The off-diagonal elements $\theta_{kl}(t)$ are  arbitrary provided that $\theta_{kl}(t)$ is positive definite. The simplest choice corresponds to $\theta_{kl}(t)=0$ for $k \neq l$. It finally gives
\begin{equation}\label{}
    \Phi_t \rho = \Theta_t N_t^{-1} \rho = \sum_{k=1}^N \frac{\dot{n}_{kk}(t)}{n_{kk}(t)} \, e_{kk} \, \rho\,  e_{kk}\ ,
\end{equation}
and hence
\begin{equation}\label{}
L_t \rho = (\Phi_t - \Psi_t)\rho = \sum_{k\neq l}   \frac{\dot{n}_{kl}(t)}{n_{kl}(t)}  \, e_{kk} \, \rho\,  e_{ll}\ ,
\end{equation}
provides the pure decoherence dynamics: $\Lambda_t \rho = \sum_{k,l=1}^N c_{kl}(t)  \, e_{kk} \, \rho\,  e_{ll}\,$ with
\begin{equation}\label{}
    c_{kl}(t) = n_{kl}(t) \ , \ (k\neq l) \ \ \ \mbox{and}\ \ \ c_{kk}=1\ .
\end{equation}
The matrix $c_{kl}(t)$ is by construction positive definite.
\end{Example}

\section{Characterizing Markovian dynamics}  \label{CRI}

In this section we analyze special properties of divisible (and hence Markovian) dynamical maps.
Let us recall, that if a linear map $\Lambda : \mathcal{T}(\mathcal{H}) \ra \mathcal{T}(\mathcal{H})$ is trace preserving, then $\Lambda$ is positive if and only if
\begin{equation}\label{contr}
    || \Lambda a ||_1 \leq ||a||_1\ ,
\end{equation}
for all hermitian $a$. Note, that $\Lambda$ needs not be contractive for non-hermitian elements. However, if $\Lambda$ is completely positive, then (\ref{contr}) holds for all $a \in \mathcal{B}(\mathcal{H})$. Actually, it turns out \cite{Szarek}  that  if $\Lambda$ is 2-positive and trace preserving, then
\begin{equation}\label{}
    ||\Lambda||_1 := \sup_{||a||_1=1} ||\Lambda a||_1 = \sup_{||a||_1=1; a^\dagger =a} ||\Lambda a||_1 = 1\ .
\end{equation}

\begin{cor} If $\Lambda_t$ is a dynamical map, then $||\Lambda_t||_1=1$, that is, $\Lambda_t$ is contractive in the trace norm
\begin{equation}\label{}
    || \Lambda_t a ||_1 \leq ||a||_1\ .
\end{equation}
Moreover, if $\Lambda_t$ is a divisible map, then
\begin{equation}\label{}
    \frac{d}{dt}\, || \Lambda_t a ||_1 \leq 0 \ ,
\end{equation}
for an arbitrary $a \in \mathcal{T}(\mathcal{H})$.
\end{cor}
Similar property holds for dynamical maps in the Heisenberg picture. Recall, that if $\Lambda_t$ is a dynamical map in the Schr\"odinger picture, then its dual  $\Lambda_t^\# : \BH \ra \BH$ corresponds to the Heisenberg picture. It is clear that $\Lambda_t^\#$ is a unital completely positive map for all $t \geq 0$. If $\Lambda^\#$ is unital and completely positive, then its operator norm satisfies
\begin{equation}\label{}
    ||\Lambda^\#|| := \sup_{||a||=1} ||\Lambda^\# a||  = 1\ .
\end{equation}

\begin{cor} If $\Lambda_t^\#$ is a dynamical map in the Heisenberg picture, then $||\Lambda_t^\#||=1$, that is, $\Lambda_t^\#$ is contractive in the operator norm
\begin{equation}\label{}
    || \Lambda_t^\# a || \leq ||a||\ .
\end{equation}
Moreover, if $\Lambda_t$ is a divisible map, then
\begin{equation}\label{}
    \frac{d}{dt}\, || \Lambda_t^\# a || \leq 0 \ ,
\end{equation}
for an arbitrary $a \in \mathcal{B}(\mathcal{H})$.
\end{cor}
\begin{Example} Consider once more the generator defined in (\ref{L-2}).  One has
\begin{equation}\label{}
    || \Lambda^\#_t \sigma_+ || = || e^{i\Omega_t - \Gamma_t} \sigma_+ || = e^{- \Gamma_t} ||\sigma_+|| =  e^{- \Gamma_t} \ ,
\end{equation}
where $\sigma_+ = |0\>\<1|$. It implies
\begin{equation}\label{}
    \frac{d}{dt}\, || \Lambda_t^\# \sigma_+ || = - \dot{\Gamma}_t = - \gamma_t \Gamma_t\ ,
\end{equation}
which shows that Markovianity of $\Lambda_t$ implies $\gamma_t \geq 0$.
\end{Example}
Let us observe that if the total Hamiltonian $\mathbf{H}$ of the ``system + reservoir' has a discrete spectrum
\begin{equation}\label{}
    \mathbf{H} = \sum_\alpha \epsilon_\alpha P_\alpha \ ,
\end{equation}
then dynamical map $\Lambda_t$ defined by
\begin{equation}\label{}
    \Lambda_t\rho = {\rm Tr}_R \Big[ e^{-i\mathbf{H}t} (\rho \ot \omega_R) e^{i\mathbf{H}t} \Big]\ ,
\end{equation}
has the following form
\begin{equation}\label{}
    \Lambda_t\rho = \sum_{\alpha,\beta} e^{-i(\epsilon_\alpha-\epsilon_\beta)t} \Lambda_{\alpha\beta}\rho\ ,
\end{equation}
where $\Lambda_{\alpha\beta} \rho = {\rm Tr}_R (P_\alpha [\rho \ot \omega_R] P_\beta)$. It is clear that due to the presence of the oscillatory terms $e^{-i(\epsilon_\alpha-\epsilon_\beta)t} $ the corresponding trace norm $|| \Lambda_t a ||_1$ is an almost quasi-periodic function and hence can not be monotonically decreasing. It proves that in such a case one obtains genuine non-Markovian dynamics.

If $\Lambda : \mathcal{T}(\mathcal{H}) \ra \mathcal{T}(\mathcal{H})$ is a linear map one defines so called diamond norm
\begin{equation}\label{}
    ||\Lambda||_\diamond := \sup_{||W||_1=1} ||(\oper \ot \Lambda)W||_1 \ .
\end{equation}
\begin{thm}
Let $\Lambda_t$ be a dynamical map. The following conditions are equivalent

\begin{enumerate}

\item $\Lambda_t$ is divisible,

\item $||V_{t,s}||_\diamond=1$ for all $t \geq s$,

\item one has

\begin{equation}\label{}
    \frac{d}{dt}\, ||(\oper \ot \Lambda_t)W||_1 \leq 0 \ ,
\end{equation}
for all Hermitian $W \in \mathcal{T}(\mathcal{H} \ot \mathcal{H})$.

\end{enumerate}

\end{thm}
The corresponding theorem in the Heisenberg picture may be formulated as follows: recall that $\Lambda^\# : \BH \ra \BH$ is completely bounded if
\begin{equation}\label{}
    ||\Lambda^\#||_{\rm cb} := ||\oper \ot \Lambda^\#|| < \infty\ ,
\end{equation}
and it is completely contractive if $||\Lambda^\#||_{\rm cb} \leq 1$.

\begin{thm}
Let $\Lambda^\#_t$ be a dynamical map in the Heisenberg picture. The following conditions are equivalent

\begin{enumerate}

\item $\Lambda_t^\#$ is divisible,

\item $||V_{t,s}^\#||_{\rm cb}=1$ for all $t \geq s$, and hence $V_{t,s}^\#$ is completely contractive,

\item one has

\begin{equation}\label{}
    \frac{d}{dt}\, ||(\oper \ot \Lambda_t^\#)A|| \leq 0 \ ,
\end{equation}
for all  $A \in \mathcal{B}(\mathcal{H} \ot \mathcal{H})$.

\end{enumerate}

\end{thm}
The Markovian evolution may be characterized in a slightly different way: we know that $\Lambda_t$ corresponds to Markovian evolution iff the 2-parameter family of propagators $V_{t,s}$ is completely positive for $t \geq s$. Denote by $\psi^+$ maximally entangled state in $\mathcal{H} \ot \mathcal{H}$ and let $P^+ = |\psi^+\>\<\psi^+|$. Note that $V_{t,s}$ is completely positive if and only if $(\oper \ot V_{t,s})P^+ \geq 0$ which is equivalent to the following simple condition
\begin{equation}\label{}
    v_{t,s} := ||(\oper \ot V_{t,s})P^+||_1 = 1\ .
\end{equation}
Let us define
\begin{equation}\label{}
    g_t := \frac{d v_{u,t}}{du}\, \Big|_{u=t} \ ,
\end{equation}
that is,
\begin{equation}\label{}
    g_t  =  \lim_{\epsilon \ra 0+} \frac{v_{t+\epsilon,t} - 1}{\epsilon}\ ,
\end{equation}
where we have used $v_{t,t}=1$. Taking into account that
\begin{equation}\label{}
    V_{t+\epsilon,t} = \oper + \epsilon L_t + O(\epsilon^2)\ ,
\end{equation}
one finds
\begin{equation}\label{}
    g_t = \lim_{\epsilon \ra 0+} \frac{|| P^+ + \epsilon (\oper \ot L_t)P^+||_1 - 1}{\epsilon}\ .
\end{equation}

\begin{cor}[\cite{RHP}]
A map $\Lambda_t$ is divisible if and only if $g_t=0$ for all $t\geq 0$.
\end{cor}

Finally, let us provide characterization of the corresponding generator in the Heisenberg picture.
Let us recall that if $\Lambda^\# : \BH \ra \BH$ is unital and 2-positive the following Kadison inequality holds
\begin{equation}\label{}
    \Lambda^\#(a a^*) \geq \Lambda^\#(a) \Lambda^\#(a^*)\ .
\end{equation}
This inequality may be used to characterize Markovian generators. Note, that Markovian dynamics $\Lambda_t^\#$ satisfies
\begin{equation}\label{}
    \partial_t V_{t,s}^\# = V_{t,s}^\# L_t^\#\ , \ \ \ V_{s,s}^\#=\oper\ ,
\end{equation}
where $V_{t,s}^\#$ denotes the dual propagator. Now, differentiating the Kadison inequality
\begin{equation}\label{}
 V_{t,s}^\#(a a^*) \geq  V_{t,s}^\#(a)  V_{t,s}^\#(a^*)\ ,
\end{equation}
one finds
\begin{equation}\label{}
    V_{t,s}^\#L_t^\#(a a^*)   \geq V_{t,s}^\# L_t^\# (a)\cdot V_{t,s}^\#(a^*) + V_{t,s}^\#(a)\cdot  V_{t,s}^\# L_t^\#(a^*)\ .
\end{equation}
Taking $t=s$ and using $V_{t,t}^\#=\oper\,$  one gets
\begin{equation}\label{}
    L_t^\#(a a^*)   \geq L_t^\# (a)\cdot a^* + a \cdot L_t^\#(a^*)\ .
\end{equation}

\begin{definition}
A hermitian map $\Psi : \BH \, \ra\,  \BH$ is dissipative iff
$$  \Psi(a a^*)   \geq \Psi(a)\cdot a^* + a \cdot \Psi(a^*)\ , $$
for all $a \in \BH$. $\Psi$ is completely dissipative if $\oper \ot \Psi$ is dissipative.
\end{definition}
One has
\begin{proposition}
$\Lambda_t^\#$ defines Markovian dynamics in the Heisenberg picture if and only if $L_t^\#$ is completely dissipative.
\end{proposition}

\section{Simple criteria for non-Markovianity} \label{CRI}

In this section we develop a series of criteria for non-Markovian dynamics. It turns out that dynamics represented by a divisible map displays characteristic monotonic behavior for several interesting quantities. Breaking monotonicity reveals non-Markovian character of the corresponding quantum evolution.

\subsection{Distinguishability}

Trace norm defines a natural distance between quantum states represented by density operators: given two density operators $\rho$ and $\sigma$ one defines
\begin{equation}\label{}
    D[\rho,\sigma] = \frac 12 ||\rho-\sigma||_1\ .
\end{equation}
The quantity $D[\rho,\sigma]$ is usually interpreted as a measure of distinguishability of quantum states $\rho$ and $\sigma$. It is well known that if $\Lambda$ is a positive trace-preserving map, then
\begin{equation}\label{}
    D[\Lambda\rho,\Lambda\sigma] \leq  D[\rho,\sigma]\ .
\end{equation}

\begin{cor}
If $\Lambda_t$ is a divisible map, then
\begin{equation}\label{DM}
   \frac{d}{dt}\,  D[\Lambda_t\rho,\Lambda_t\sigma] \leq  0\ ,
\end{equation}
that is, for the Markovian evolution distinguishability of any pair of initial states monotonically decreases.
\end{cor}
It is well known that if $\Lambda_t$ corresponds to the unitary dynamics $\Lambda_t\rho = U_t \rho U_t^\dagger$, with unitary $U_t$, then $D[\Lambda_t\rho,\Lambda_t\sigma] = 0$. Moreover, if $\Lambda_t = e^{Lt}$ represents dynamical semigroup, then $D[\Lambda_t\rho,\Lambda_t\sigma] <0$.
The above property was used by Breuer et al \cite{BLP} as another definition of Markovianity.
This criterion identifies non-Markovian dynamics with certain  physical features of the
system-reservoir interaction. They define non-Markovian dynamics as
a time evolution for the open system characterized by a temporary
flow of information from the environment back into the system. This
backflow of information may manifest itself as an increase in the
distinguishability of pairs of evolving quantum states. It turns out that these two concepts of Markovianity do not agree (see e.g. \cite{Angel,M}). Clearly, divisibility implies (\ref{DM}) but the converse needs not be true.

\subsection{Fidelity}

Given two density operators $\rho$ and $\sigma$ one defines Uhlmann fidelity
\begin{equation}\label{}
    F(\rho,\sigma) = \Big({\rm Tr}\, \Big[ \sqrt{\sqrt{\rho}\, \sigma\, \sqrt{\rho}} \Big] \Big)^2\ .
\end{equation}
Equivalently, one has
\begin{equation}\label{}
    F(\rho,\sigma) = || \sqrt{\rho} \sqrt{\sigma} ||_1^2 \ ,
\end{equation}
which shows that $F(\rho,\sigma) = F(\sigma,\rho)$.
One proves
\begin{equation}\label{}
    1 - F(\rho,\sigma) \leq D[\rho,\sigma] \leq \sqrt{1-F(\rho,\sigma)^2}\ .
\end{equation}
Moreover, If $\Lambda$ is a quantum channel, then
\begin{equation}\label{}
    F(\rho,\sigma) \leq  F(\Lambda\rho,\Lambda\sigma)\ .
\end{equation}

\begin{cor}
If $\Lambda_t$ is a divisible map, then
\begin{equation}\label{}
    \frac{d}{dt} F(\Lambda_t\rho,\Lambda_t\sigma) \geq 0 \ .
\end{equation}
\end{cor}

\subsection{Entropic quantities}

Let us recall the definition of Renyi $S_\alpha$ and Tsallis $T_q$ relative entropies
\begin{equation}\label{}
    S_\alpha(\rho\, ||\, \sigma) = \frac{1}{\alpha-1} \log\Big[ {\rm Tr}\,\rho^\alpha \sigma^{1-\alpha} \Big] \ ,
\end{equation}
for $\alpha \in [0,1) \cup (1,\infty)$, and
\begin{equation}\label{}
    T_q(\rho\, ||\, \sigma) = \frac{1}{1-q}\, \Big[ 1 - {\rm Tr} \,\rho^q \sigma^{1-q} \Big] \ ,
\end{equation}
for $q\in [0,1)$. Note, that in the limit
$$ \lim_{\alpha \ra 1} S_\alpha(\rho\, ||\, \sigma) =  \lim_{q \ra 1} T_q(\rho\, ||\, \sigma) = S(\rho\, ||\, \sigma)\ , $$
one recovers well known formula for relative entropy
\begin{equation}\label{}
    S(\rho\,||\,\sigma) = {\rm Tr}(\rho [\log \rho - \log \sigma] )\ .
\end{equation}
It turns out \cite{QIT,Petz} that if $\Lambda$ is a quantum channel then $S_\alpha$ and $T_q$ satisfy
\begin{equation}\label{}
    S_\alpha(\Lambda\rho\, ||\, \Lambda\sigma) \leq  S_\alpha(\rho\, ||\, \sigma)\ ,
    \ \ \ \ T_q(\Lambda\rho\, ||\, \Lambda\sigma) \leq  T_q(\rho\, ||\, \sigma)\ ,
\end{equation}
for $\alpha \in [0,1) \cup (1,2]$ and $q \in [0,1)$. Clearly, the same property holds for the relative entropy.
\begin{cor} If $\Lambda_t$ is a divisible map, then
\begin{equation}\label{}
    \frac{d}{dt}\, S_\alpha(\Lambda_t\rho\, ||\, \Lambda_t\sigma)  \leq 0 \ , \ \ \ \
    \frac{d}{dt}\, T_q(\Lambda_t\rho\, ||\, \Lambda_t\sigma)  \leq 0 \ ,
\end{equation}
for $\alpha \in [0,1) \cup (1,2]$, $q\in [0,1)$, and
\begin{equation}\label{}
    \frac{d}{dt}\, S(\Lambda_t\rho\, ||\, \Lambda_t\sigma)  \leq 0 \ .
\end{equation}
\end{cor}

\subsection{Entanglement measures}

Consider a composed system living in $\mathcal{H} \ot \mathcal{H}'$ and let $W$ be an arbitrary density matrix in $\mathcal{H} \ot \mathcal{H}'$. It is well known \cite{QIT} that for arbitrary genuine entanglement measure $\mathcal{E}$ one has
\begin{equation}\label{}
    \mathcal{E}([\Phi \ot \Phi'] W) \leq \mathcal{E}(W)\ ,
\end{equation}
where $\Phi : \TTH \ra \TTH$ and $\Phi' : \mathcal{T}(\mathcal{H}') \ra \mathcal{T}(\mathcal{H}')$ are quantum channels.
Denote by $W_t$ the trajectory   $\, W_t = (\Lambda_t \ot \oper)W\,$ starting at $W$. Now, if $\mathcal{E}$ is an entanglement measure, then
\begin{equation}\label{}
    \mathcal{E}(W_t) \leq \mathcal{E}(W)\ ,
\end{equation}
for an arbitrary dynamical map $\Lambda_t$. It is, therefore, clear that if $\Lambda_t$ is divisible then
\begin{equation}\label{}
    \frac{d}{dt} \, \mathcal{E}(W_t) \leq 0 \ ,
\end{equation}
for each initial state $W$. Of course the above relation is nontrivial only if the initial state $W$ is entangled. In particular if $W= P^+$ (maximally entangled state), then entanglement measured by $\cal E$ monotonically decreases from the maximal value ${\cal E}(P^+)$ \cite{RHP}.

\section{Conclusions}   \label{CON}

We characterized a class of Markovian dynamics using the concept of divisible dynamical map.
Interestingly Markovian dynamics is fully controlled in the local approach {\em via} the properties of the corresponding local generator. Characterization of Markovianity in terms of the memory kernel $\mathcal{K}_t$ is an open problem. It should be stressed that the standard Markovian master equation (\ref{M-0}) is defined by the corresponding macroscopic model and a suitable Markovian approximation. Note, that for general Markovian evolution characterized by time-dependent local generator the construction of the corresponding macroscopic model is not known.

Moreover we provided a family of criteria which can distinguish Markovian and non-Markovian dynamics. These {\em Markovianity criteria} are based on a simple observation that Markovian dynamics implies monotonic behavior of several well known quantities like distinguishability of states, fidelity, relative entropy and genuine entanglement measures.

We stress that the problem of characterization of admissible $L_t$ and $\mathcal{K}_t$ giving rise to legitimate dynamical map $\Lambda_t$ is rather untractable. Only commutative case is fully controlled. One may wonder if there is another way to describe dynamics of an open system.
In a forthcoming paper we propose a new approach to this problem.

\end{document}